\documentclass[onecolumn,prd,aps,showpacs,11pt]{revtex4}
\usepackage{epsf,epsfig,graphicx,amsmath,amssymb}

  \newcount\hour \newcount\minute
  \hour=\time \divide \hour by 60
  \minute=\time
  \newcommand{\mydate}{\ \today \ - \number\hour :\ifnum \minute<10 0\fi
\number\minute}

\def\be{\begin{eqnarray}}
\def\en{\end{eqnarray}}
\def\non{\nonumber}

\def\3bar{{\bf \bar 3}}
\def\6bar{{\bf \bar 6}}
\def\10bar{{\bf \ov{10}}}
\def\ov{\overline}

\def\pr{{Phys. Rev.}~}
\def\prl{{ Phys. Rev. Lett.}~}
\def\pl{{ Phys. Lett.}~}
\def\np{{ Nucl. Phys.}~}

\def\lsim{ {\ \lower-1.2pt\vbox{\hbox{\rlap{$<$}\lower5pt\vbox{\hbox{$\sim$}
}}}\ } }
\def\gsim{ {\ \lower-1.2pt\vbox{\hbox{\rlap{$>$}\lower5pt\vbox{\hbox{$\sim$}
}}}\ } }

\begin{document}
\title{Role of electromagnetic dipole operator in the electroweak penguin dominated vector meson decays of $B$ meson}
\thanks{This
work is partly supported by National Science Foundation of China
under contract No. 10475085}

\author{ Cai-Dian L\"u$^{a,b}$,Yue-Long Shen$^b$ and
Wei Wang$^b$} \affiliation{\it \small
 a CCAST (World Laboratory), P.O. Box 8730,
   Beijing 100080, P.R. China\\
\it \small  b Institute of High Energy Physics, CAS, P.O.Box
918(4), \it \small 100049, P.R. China\footnote {Mailing address}}

 \begin{abstract}
 The pure annihilation type decays $B^0_d\to\phi\gamma$ and $B_s\to\rho\gamma$  receive only
 color suppressed penguin contributions with a very small branching
 ratio in the standard model. When we include  the previously neglected electromagnetic dipole
operator, the branching ratios can be enhanced one order magnitude
larger than previous study using QCD factorization approach. That
is ${\cal BR}(\bar B^0_d\to\phi\gamma)\simeq 1 \times 10^{-11}$
and ${\cal BR}(B_s\to\rho\gamma) \sim (6-16)\times 10^{-10}$. The
new effect can also give a large contribution, of order $10^{-9}$,
to transverse polarization of $B\to\phi\rho$ and $B\to \omega\phi$
which is comparable to the longitudinal part. These effects can be
detected in the LHCb experiment and the Super-B factories.
\end{abstract}
\maketitle

The charmless  two-body B decays provide  rigorous information on
flavor changing phenomena in the weak interactions of quarks, so
it has attracted much attention to check the consistency of the
standard model (SM) and to explore the existence of the possible
new physics beyond standard model. On the other hand, the rare B
decays also serve as a laboratory for hadronic dynamics. However,
predictions for many interesting decays are always polluted by the
hadronization, which have hindered us in extracting weak
interaction information precisely from the available measurements.
To handle the decay processes, many factorization approaches have
been developed to separate nonperturbative dynamics from the
perturbative part, such as naive factorization \cite{Naive},
generalized factorization \cite{AKL, CCTY}, the QCD factorization
(QCDF) \cite{QCDF}, perturbative QCD approach (PQCD) \cite{PQCD}
and the soft-collinear effective theory (SCET) \cite{SCET}.

\begin{figure}
\scalebox{1}[1]{\includegraphics{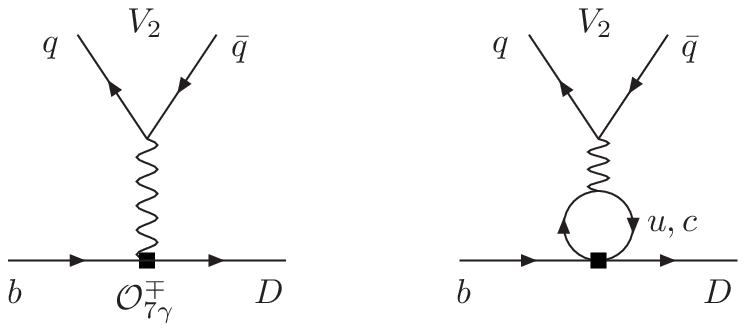}}
\caption{\label{Q7diags}New contributions to $B \to VV$ decays}
\end{figure}

Two vector decays of B meson  shed light on the helicity structure
of weak interactions through polarization studies. The naive
counting rules \cite{Kagan} based on the factorization approaches
predict that the longitudinal polarization dominates the decay
ratios and the transverse polarizations are suppressed due to the
helicity flips of the quark in the final state hadrons. But for
penguin-dominated decays such as $B\to\phi K^*$ and $ B\to\rho K^*$
\cite{rhophi}, large transverse polarizations are observed. Beneke,
Rohrer and Yang \cite{BRY} discussed a novel electromagnetic penguin
contribution to the transverse helicity amplitudes, which  is shown
 in Fig.\ref{Q7diags}. The neutral vector meson generated by the
photon is mainly transversely polarized. Though suppressed by the
fine structure constant $\alpha_{em}$, it is enhanced over the
leading terms by
 $m_B/\Lambda$. So the transverse helicity amplitude of $B\to \rho^0
 K^*$ can be enhanced by this effect. The validity of this effect can be
tested in the ratio of transverse polarization for $B\to \rho^0
 K^*$ over $B\to \rho^+ K^*$ in the near future.

\begin{figure}
\scalebox{1}[1]{\includegraphics{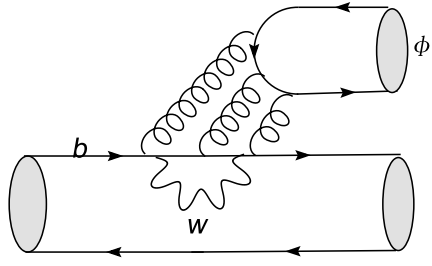}}
\caption{\label{hairpin}Hairpin diagrams for $B \to \phi\rho$ and
$B\to\phi\omega$ processes, and similar for $B\to\phi\gamma$.}
\end{figure}

It may play an important role in the electroweak penguin dominated
vector meson decays of $B$ meson. In the standard model, it is
suppressed to produce a $\phi$ meson through strong interaction
for three or more gluons are needed which are called hairpin
diagrams \cite{xingzz} as shown in Fig. \ref{hairpin}.  So the
electromagnetic dipole operator may become important in
$B\to\phi\gamma$ and $B\to\phi\rho$ decays. We examine the new
effect in these modes below.

In the standard model, the effective weak Hamiltonian mediating
flavor-changing neutral current transitions of the type $b\to D
(D=d,s)$ has the form \cite{Buras}:
 \be
 {\cal H}_{eff}=&&{G_F\over \sqrt 2}\Big[\sum\limits_{p=u,c}V_{pb}V^*_{pD}\Big(
 C_1O_1^p+C_2O_2^p\Big)-V_{tb}V^*_{tD}\sum\limits_{i=3}^{10}\Big(
 C_iO_i+C_{7\gamma}O_{7\gamma}+C_{8g}O_{8g}\Big)\Big].
 \en
 For convenience, we specify the below operators in ${\cal
 H}_{eff} $:
 \be
O^p_1&=&\bar p_\beta \gamma^\mu L b_\alpha \cdot \bar D_\alpha
 \gamma_\mu Lp_\beta,\,\,\,\,\,\,\,\,
 O^p_2=\bar p_\alpha \gamma^\mu L b_\alpha \cdot \bar D_\beta
 \gamma_\mu Lp_\beta,\non\\
 O_3&=&\bar D_\alpha \gamma^\mu L b_\alpha \cdot \sum_{q}\bar q_\beta
 \gamma_\mu Lq_\beta,\,\,\,\,\,\,
 O_4=\bar D_\beta \gamma^\mu L b_\alpha \cdot\sum_{q} \bar q_\alpha
 \gamma_\mu Lq_\beta,\non\\
  O_5&=&\bar D_\alpha \gamma^\mu L b_\alpha \cdot \sum_{q}\bar
  q_\beta\gamma_\mu Rp_\beta,\,\,\,
 O_6=\bar D_\beta \gamma^\mu L b_\alpha \cdot\sum_{q} \bar q_\alpha
 \gamma_\mu Rq_\beta,\non\\
  O_7&=&\frac{3}{2}\bar D_\alpha \gamma^\mu L b_\alpha \cdot\sum_{q}e_q \bar q_\alpha
 \gamma_\mu Rq_\beta,\,\,\,
 O_8=\frac{3}{2}\bar D_\beta \gamma^\mu L b_\alpha \cdot\sum_{q}e_q \bar q_\alpha
 \gamma_\mu Rq_\beta,\non\\
  O_9&=&\frac{3}{2}\bar D_\alpha \gamma^\mu L b_\alpha \cdot\sum_{q}e_q \bar
  q_\beta\gamma_\mu Lq_\beta,\,\,\,
 O_{10}=\frac{3}{2}\bar D_\beta \gamma^\mu L b_\alpha \cdot\sum_{q}e_q \bar q_\alpha
 \gamma_\mu Lq_\beta,\non\\
  O_{7\gamma}&=&-{em_b\over 8\pi^2}\bar
  D\sigma^{\mu\nu}F_{\mu\nu}(1+\gamma_5)b,
\en where $\alpha$ and $\beta$ are the SU(3) color indices and L
and R are the left- and right-handed projections operator with
$L=(1-\gamma_5)$, $R=(1+\gamma_5)$. The sum over q runs over the
quark fields that are active at the scale $\mu={\cal O}(m_b)$,
i.e., $q=u,d,s,c,b$. The Wilson coefficients evaluated at
$\mu=\bar m_b(m_b)=4.4GeV$ from $m_t=170 GeV$ in NDR with
$\Lambda^{(5)}_{\bar{MS}}=225MeV$ ($C_{7\gamma}$  evaluated at
$\mu=5.0$ GeV from $m_t=170$ GeV with $\alpha_s^{(5)}(M_Z)=0.118$)
are: \be
C_1&=&-0.185,\,\,C_2=1.082,\,\,C_3=0.014,\,\,C_4=-0.035,\,\,C_5=0.009\,\,C_6=-0.041,\non\\
C_7&=&-0.002/137,\,\,C_8=0.054/137,\,\,C_9=-1.292/137,\,\,C_{10}=0.263/137,
C_{7\gamma}=0.300.\label{Wilson}
\en

\begin{figure}
\scalebox{1}[1]{\includegraphics{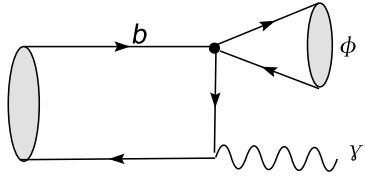}}
\caption{\label{fourquark}The leading diagram from four-quark
operators for $\bar B^0\to\phi\gamma$ and $B_s\to\rho\gamma$. }
\end{figure}

\begin{figure}
\scalebox{1}[1]{\includegraphics{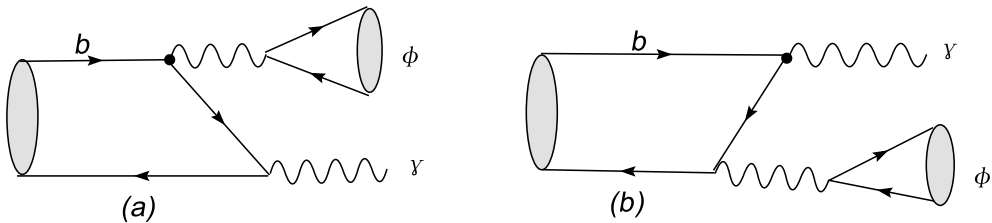}}
\caption{\label{enhancementphigamma}The electromagnetic dipole
operator (denoted by heavy dot) contribution to $\bar
B^0\to\phi\gamma$ and $B_s\to\rho\gamma$. }
\end{figure}

The radiative decay $\bar B^0_d\to\phi\gamma$ has been studied
using QCD factorization approach in \cite{Y.D.Yang} and using
perturbative QCD approach in \cite{LL}. There are four diagrams
contributing to this decay mode from four-quark operators. The two
diagrams with photon emitting from $s$ and $\bar s$ cancel with
each other. The diagram with photon emitting from the $b$ quark is
suppressed by $1/m_b$ compared to the contribution with photon
emitting from the light quark in $B$ meson. The dominant
contribution from four-quark operators to $\bar
B^0_d\to\phi\gamma$ is illustrated in Fig. \ref{fourquark}. In
naive factorization, this process are proportional to
$C_3+C_4/3+C_5+C_6/3-0.5(C_7+C_8/3+C_9+C_{10}/3)$. They have very
tiny branching ratios  due to the large cancelations between
Wilson coefficients: $C_3\simeq-C_4/3$ and $C_5\simeq-C_6/3$. This
cancellation implies the fact that the dominant diagrams for these
decays are the three-gluon hairpin diagrams. To be even worse, the
QCD penguins cancel with the electroweak penguins:
$C_3+C_4/3+C_5+C_6/3=-0.002$,
$-0.5(C_7+C_8/3+C_9+C_{10}/3)=0.004$.  It is dominated by the
photon radiating from the light quark in B meson. Because the
$\phi$ meson is transversely polarized, so the amplitude is
suppressed by a power of $\Lambda_{QCD}/m_B$ further.
Factorization properties of the radiative decays $B\to V\gamma$
have been analyzed  at the leading power in $1/m_b$ in Ref.
\cite{BHN} (see also \cite{CK}) using the soft-collinear effective
theory. None of the leading power operators in SCET$_{II}$ will
contribute to $\bar B^0_d\to\phi\gamma$, we have to perform the
subleading power analysis and this also implies that this mode is
suppressed. In \cite{Y.D.Yang}, the branching ratios of naive
factorization and QCD factorization which includes the
non-factorizable contributions are:  \be {\cal BR}(\bar B^0_d\to
\phi\gamma)_{NF}=3.5\times 10^{-13},\\
 {\cal BR}(\bar B^0_d\to
\phi\gamma)_{QCDF}=3.6\times 10^{-12}.\en The above results  are
too rare to be measured in the running B factories BaBar and Belle
and may be a probe to detect the new physics. As an example, it
can be treated as a probe of R-parity violating couplings (RPV).

In the effective theory, when matching QCD to SCET$_I$ at the
leading power, the electromagnetic dipole operators can be
factorized into (we use the notation of \cite{BHN}): \be
J^A&=&\bar {\cal X }_{hc}{\cal A}^{em}_{{\bar {hc}}\perp}\Gamma^\prime h_s,\non\\
J^D&=&\bar {\cal X }_{\bar {hc}}{\cal A}^{em}_{{
{hc}}\perp}\Gamma^\prime h_s.\en Including QED in SCET, there is a
collinear photon field with unsuppressed interactions with
collinear quarks of the same direction. With the additional
operator ${\cal A}$ overlaps with the transversely polarized
vector meson, the $J^A$ and $J^D$ can be matched onto the same
operator in SCET$_{II}$: \be O=\bar {\cal Q }_{s}{\cal A}^{em}_{{
{c}}\perp}{\cal A}^{em}_{{\bar {c}}\perp}\Gamma^\prime {\cal H}_s.
\en At tree level, these operators correspond to the two diagrams
in Fig.\ref{enhancementphigamma}. Using the naive factorization
approach the decay amplitude reads:
 \be A(\bar
B^0_d\to\phi\gamma)_a&=&\frac{G_F}{\sqrt{2}}V_{tb}^*V_{td}\frac{-2\bar{m_b}e^3f_\phi
Q_dQ_sN_c}{\pi^2m_\phi m_B\sqrt{2N_c}}C_{7\gamma}\lambda_{B}^{-1}
\times\epsilon^{*}_{\gamma\alpha}p_{\gamma
\beta}p_{\phi\mu}\epsilon^*_{\phi\nu}(g^{\alpha\nu}g^{\beta\mu}
-g^{\alpha\mu}g^{\beta\nu}-i\epsilon^{\alpha\beta\mu\nu}),\en
where $\epsilon_\gamma,p_\gamma$ and $\epsilon_\phi,p_\phi$ are
the momentum and polarization for the photon and $\phi$ meson.
$Q_d=Q_s=-1/3, N_c=3$. $\lambda_B^{-1}$ is the first inverse
moment of the B meson's wave function and defined as
 \be
\lambda^{-1}_B=\int^1_0dx\frac{\phi_B(x)}{x}, \en where $x$ is the
momentum fraction carried by the light quark in the B meson. The
neutral vector meson is generated by the intermediate photon, so
we can view  the two diagrams in Fig. \ref{enhancementphigamma} as
$B_{d,s}\to\gamma\gamma$ \cite{gammagamma} at first. The second
diagram can be related to the first one by the crossing symmetry,
so it gives the same contribution as the first one.

\begin{table}\caption{ Branching ratios of  $\bar
B^0_d\to\phi\gamma$ (in units of $10^{-11}$) and $B_s\to
\rho\gamma$ (in units of $10^{-10}$) using different
$\lambda^{-1}_{B_{(s)}}$}
\begin{tabular}{c|c|c|c}
 \hline \hline
  $\lambda_B^{-1}$&0.4&0.5&0.6\\
\hline$\bar B^0_d\to\phi\gamma$(without $O_{7\gamma}$) & $0.04 $
& $0.06$ & $0.09$ \\
$\bar B^0_d\to\phi\gamma$(with $O_{7\gamma}$) & $0.86 $
& $1.3$ & $1.9$ \\
\hline\hline
 $\lambda_{B_s}^{-1}$&0.4&0.5&0.6\\
 \hline
$B_s\to \rho\gamma$ (without $O_{7\gamma}$) &$0.45$ & $0.69$& $1.0$ \\
$B_s\to \rho\gamma$(with $O_{7\gamma}$) &$5.9$& $9.2$& $13$\\
\hline\hline
\end{tabular}\label{t1}
\end{table}

There are many candidates for B meson distribution
 amplitudes \cite{PQCD,wavefunction}, but here only
 $\lambda_B^{-1}$ is involved and we use three values for this
 parameter:  $\lambda_B^{-1}=0.4,0.5,0.6$.
  We adopt other parameters
the same as \cite{Y.D.Yang,LL}, then numerical results for the
branching ratios of these two modes are given in Table \ref{t1}.
For $B_s\to\rho\gamma$, there is also a contribution from the tree
operators $O_1$ and $O_2$. But this contribution is suppressed by
the CKM matrix elements and roughly is of order $10^{-12}$, so we
neglect it in the Table. We find that the branching ratios are
enhanced sizably by the new contribution relative to the result
from QCD factorization approach. Although it is hardly to be
measured at BaBar and Belle, we state that the contribution of
order $10^{-11}$ for $\bar B^0_d\to\phi\gamma$ would not be the
evidence of the activity of new physics. As for
$B_s\to\rho\gamma$, ${\cal BR}\sim(6-16)\times 10^{-10}$ which
might be detected at the LHCb experiment.

 For $B\to\rho\phi$ and $B\to
\omega\phi$, they are very similar to the above radiative decays.
These processes can also only be produced via the so-called
hair-pin diagrams
 as in Fig. \ref{hairpin} \cite{xingzz} which are highly suppressed and the branching ratios in
generalized factorization \cite{AKL} are:
\be {\cal{BR}}(B^+\to \rho^+\phi)&=&0.04\times 10^{-7},\non\\
{\cal BR}(B^0\to \rho^0\phi)&=&0.02\times 10^{-7},\non\\
{\cal BR}(B^0\to \omega\phi)&=&0.02\times 10^{-7}.\en In this
calculations, the electromagnetic dipole operator $O_{7\gamma}$ is
also neglected.  This operator contributes only to the
transversely polarized amplitude. The contribution can be
factorized into \cite{BRY}: \be {\cal A}(B\to
V_1V_2)=im_{V_2}m_B2T^{V_1}_1(0)f_{V_2}a_{V_2} \times
\Big(-\frac{2\alpha_{em}}{3\pi}\Big)C_{7\gamma}\frac{m_Bm_b}{m_{V_2}^2},\en
where $V_2$ is a neutral vector meson produced by the photon.
$a_{V_2}$ is a constant which depends on the quark-flavor
composition of $V_2$: $a_\phi=-1/2$, $a_\rho=3/(2\sqrt2)$. The
factor $\sqrt 2$ arises from the normalization of the $\rho^0$
meson. $T^{V_1}_1(0)$ is the QCD tensor form factor of $B\to V_1$,
$f_{V_2}$ and $m_{V_2}$ is the decay constant and mass of $V_2$.
The amplitude takes the given value only when both $V_1$ and $V_2$
have negative helicity, but is zero otherwise.

The four-quark operators in Fig.\ref{Q7diags}(b) from the
effective Hamiltonian could give corrections to the above
expression  and they have been computed in next-to-leading order
in the context of factorization of exclusive radiative B decays
\cite{BFS,BB}. After incorporating this contribution, the Wilson
coefficient $C_{7\gamma}$ is replaced by ${\cal C}^\prime_{7}$
\cite{BFS}. The tensor form factor $T^{V_1}_1(0)$  is a
nonperturbative parameter. In PQCD approach, it can be factorized
into the convolution of the wave function and the hard part. We
take the results from the PQCD calculations \cite{LY}: \be
T^{B\to\rho}_1(0)=0.41,\,\,\,T^{B\to\omega}_1(0)=0.38. \en Using
the above inputs and
$f_\phi=0.25\mbox{4GeV},m_\phi=1.02\mbox{GeV},m_B=5.28\mbox{GeV},|V_{td}|=0.008,|V_{tb}|=1.0$,
we get \be
{\cal BR}(B^+\to \rho^+\phi)&=&3.0^{+0.5}_{-0.5}(1.6)\times 10^{-9},\non\\
{\cal BR}(B^0\to \rho^0\phi)&=&1.4^{+0.2}_{-0.2}(0.77)\times 10^{-9},\non\\
{\cal BR}(B^0\to \omega\phi)&=&1.2^{+0.2}_{-0.2}(0.66)\times
10^{-9}.\en Results in the parentheses are performed by using the
leading order Wilson coefficient for $C_{7\gamma}$, while the
others are evaluated by using $|{\cal C}_7^\prime
|^2=0.165^{+0.018}_{-0.017}$ \cite{BFS}. From the results, we can
see that the new contribution is comparable to the original one
which may imply that the large transverse polarization in this
mode. Compared with the experimental result by BaBar collaboration
\cite{BaBarphiomega}:
 \be {\cal BR}(B^0\to
\phi\omega)=0.1\pm0.5\pm0.1(<1.2)\times
10^{-6},\label{expphiomega}\en we can see that these modes can
hardly be measured in the running B factories. If we replace $B$
by $B_s$ ($V_{td}\to V_{ts}$) and assume SU(3) symmetry to give an
estimate, the corresponding processes $B_s\to\phi\rho^0$ and
$B_s\to\phi\omega$ can be enhanced to $10^{-7}$  which may be
measured at the future LHCb experiment.

In summary, we have studied  the enhancement of  $\bar
B^0_d\to\phi\gamma$ and $B_s\to \rho\gamma$ due to the
electromagnetic dipole operators. This effect could produce
sizable contributions to  $\bar B^0_d\to\phi\gamma$ to make a
branching ratio of about $1.0\times 10^{-11}$, which is larger
than the previous study using QCD factorization approach. Although
it is hardly to be measured at BaBar and Belle, we conclude that
the contribution of order $10^{-11}$ would not be the evidence of
the activity of new physics. As for $B_s\to\rho\gamma$, the
branching ratio is $(6-16)\times 10^{-10}$. For $B\to\rho\phi$ and
$B\to \omega\phi$, the new effect also give a large contribution,
of order of $10^{-9}$, to the transverse helicity amplitude which
is comparable to the longitudinal amplitude. So after
incorporating this effect, the polarization can be changed sizably
from the naive power counting for $B\to VV$. These effects may be
detected in the future LHCb experiment.

\end{document}